\documentclass[11pt]{amsart}
\usepackage{amsmath, color}
\usepackage{amssymb}
\usepackage{amsfonts}
\usepackage{amsthm}
\usepackage{mathrsfs}
\usepackage[all]{xy}
\usepackage{cite}
\usepackage{graphicx}
\usepackage{braket}

\newcommand{\be}{\begin{equation}}
\newcommand{\ee}{\end{equation}}
\newtheorem{theorem}{Theorem}[section]
\newtheorem{lemma}[theorem]{Lemma}

\newtheorem{corollary}[theorem]{Corollary}

\theoremstyle{definition}

\theoremstyle{remark}

\numberwithin{equation}{section}

\begin{document}
\title{Super Quantum discord for general two qubit X states}
\author{Naihuan Jing, Bing Yu$^*$}
\address{Jing: School of Mathematics, South China University of Technology,
Guangzhou, Guangdong 510640, China}
\address{Department of Mathematics, North Carolina State University, Raleigh, NC 27695, USA}
\address{Yu: School of Mathematics, South China University of Technology,
Guangzhou, Guangdong 510640, China}

\thanks{{\scriptsize
\hskip -0.4 true cm MSC (2010): Primary: 81P40; Secondary: 81Qxx.
$*$Corresponding author: 806355918@qq.com}}

\maketitle

\begin{abstract}
The exact solutions of the super quantum discord are derived for general two qubit X states in terms of a one-variable function.
Several exact solutions of the super quantum discord are given for the general X-state over nontrivial regions of a seven dimensional manifold.
It is shown that the super quantum discord of the X state may increase or decreases under the phase damping channel.
\end{abstract}
\section{INTRODUCTION}
Quantum discord measures the difference between the total correlation and classical correlation based on a family of
complete mutually orthogonal projectors such as the von Neumann measurements \cite{OZ, HV}. It has been investigated in various works, and reveals new properties in quantum correlations \cite{DVB,A,GA,GGZ,CZYYO,FWBAC,H,S,L}. As quantum states are fragile to quantum measurements; when they are
undergone projective measurements, their coherence are likely to be loosed. In 1988, Aharonov, Albert, and Vaidman have proposed to use
weak measurements \cite{AAV}, which cause only small changes to the state, and it is expected that the quantum state may loose partial coherence
under the weak measurements. Recently, the quantum
discord under weak measurement, called the ``super quantum discord'' (SQD) by Singh and Pati, brings new hope for deeper insights on the quantum correlation \cite{SP}. It is known that the weak measurement captures more quantum correlation of a bipartite system than the strong (projective) measurement under certain situation. Since then, SQD has been studied in various perspectives \cite{W,LF,RS,LT}. It is known that the solution is equivalent to the optimization of a multi-variable function with seven parameters.
However, exact solutions of SQD are few for general two qubit X states, except for the case of diagonal states.

We observe that most of the previous
 methods claim that the super quantum discords are given by the entropic functions at the endpoints, which is unfortunately an incorrect statement (see counterexamples given in Example 3).
 Thus it is necessary to settle the super quantum discord in the general case of X-type states.

 The aim of this paper is to propose a brand new method to compute the super quantum discord by reducing the optimization
 to that of one-variable function. This completely solves the problem in principle. We also give
  analytical formulas of the SQD for several nontrivial regions of the parameters. To examine the dynamic behavior of
  SQD under damping channel, we also analyzed the super quantum discord through the phase damping channel.
 It is shown that the super quantum discord of the X state may decreases or increase under the damping channel.
 %as the decoherence rate increases regardless of the parameter. 
 However, there also exists an example of $X$-state
where the super quantum discord is stable or even decreasing through the whole process under the damping channel.

The article is organized as follows: in section \uppercase\expandafter{\romannumeral2}, we shall review the weak measurement formalism and the definition for super quantum discord. In section \uppercase\expandafter{\romannumeral3}, we shall give an analytic solution for the SQD of general two qubit X-type states and also show that SQD is given by the minimum of a one-variable entropy-like function.  In section \uppercase\expandafter{\romannumeral4}, we shall analyze the dynamics of super quantum discord under phase damping channel. In section \uppercase\expandafter{\romannumeral5},we shall conclude our work. Two appendixes present details proofs of lemma \ref{le:1} and theorem \ref{t:1}.

\section{THE DEFINITION FOR SUPER QUANTUM DISCORD}
Let $\Pi_0, \Pi_1$ be a pair of orthogonal projectors such that $\Pi_i\Pi_j=\delta_{ij}\Pi_i, \Pi_0+\Pi_1=I$. In order to
study more general situation, one considers the weak measurement operators which are a pair of
complete mutual parameterized orthogonal operators that are not necessarily idempotents. For any real $x\geqslant 0$, let
\begin{equation}
P(\pm x)=\sqrt{\frac{1\mp \tanh x}{2}}\Pi_0+\sqrt{\frac{1\pm\tanh x}{2}}\Pi_1.
\end{equation}
Then $P(\pm x)^2=I\pm\frac{\tanh x}2 (\Pi_1-\Pi_0)$, $P(x)P(-x)=P(x)P(-x)=0$ and $P(x)^2+P(-x)^2=I$.
Moreover, (ii) $\lim_{x\to\infty}P(x)=\Pi_1$  and  $\lim_{x\to\infty}P(-x)=\Pi_0 $.
We will call $P(x), P(-x)$ a pair of weak measurement operators
associated with $\Pi_i$ \cite{OB}.

The super quantum discord of a bipartite quantum state $\rho_{AB}$ with weak measurements on the subsystem $B$ is the difference between the quantum mutual correlation $\mathcal I(\rho_{AB})$ and the classical correlation $\mathcal J(\rho_{AB})$\cite{SP}.
Recall that
the quantum mutual information is given by \cite{P}
$$\mathcal I(\rho_{AB})=S(\rho_A)+S(\rho_B)-S(\rho_{AB}),
$$
where $S(\rho_{A}), S(\rho_{B}), S(\rho_{AB})$ are the von Neumann entropies of the reduced state $\rho_A=\mathrm{Tr}_B(\rho_{AB})$, $\rho_B=\mathrm{Tr}_A(\rho_{AB})$, and the total state $\rho_{AB}$ respectively. The classical correlation represents the information gained about the subsystem $A$ after performing the measurements ${P^B(x)}=P(x)$ on subsystem $B$ \cite{HV} and it is defined as by the supremum
\begin{align}\label{c}
\mathcal J(\rho_{AB})=S(\rho_A)-\min_{\{P^B(x)\}}S(A|\{P^B(x)\}),
\end{align}
where
\begin{align*}
S(A|P^B(x))&=p(x)S(\rho_{A|P^B(x)})+p(-x)S(\rho_{A|P^B(-x)}),\\
p(\pm x)&=\mathrm{tr}[(I_A\otimes P^B(\pm x))\rho_{AB}(I_A\otimes P^B(\pm x))],\\
\rho_{A|P^B(\pm x)}&=\frac1{p(\pm x)}\mathrm{tr}_B[(I_A\otimes P^B(\pm x))\rho_{AB}(I_A\otimes P^B(\pm x))].\\
\end{align*}
Finally the super quantum discord $SD(\rho_{AB})$ is defined as the difference between $\mathcal I(\rho_{AB})$ and $\mathcal J(\rho_{AB})$,
\begin{align}\label{superd}
SD(\rho_{AB})&=\mathcal I(\rho_{AB})-\mathcal J(\rho_{AB})\notag\\
&=S(\rho_B)-S(\rho_{AB})+\min_{\{P^B(x)\}}S(A|\{P^B(x)\}).
\end{align}
When $\lim x\to\infty$, super quantum discord becomes the usual quantum discord under the von Neumann measurements. Therefore its computation
can be extremely challenging given that the discord is a nontrivial optimization problem over a parameterized manifold with boundary.

\section{SUPER QUANTUM DISCORD for TWO QUBIT X STATES}
We consider the general two qubit X state written in the matrix form in terms of the usual basis:
\begin{align}
\rho_{AB}=\left(
  \begin{array}{cccc}
    \rho_{11} & 0 & 0 & \rho_{14} \\
    0 & \rho_{22} & \rho_{23} & 0 \\
    0 & \rho_{32} & \rho_{33} & 0 \\
    \rho_{41} & 0 & 0 & \rho_{44} \\
  \end{array}
\right).
\end{align}
As a density matrix, the coefficients $\rho_{ij}$ are complex numbers and
satisfy the following conditions:
$\sum\limits_{i=1}^4 \rho_{ii}=1$, $\rho_{22}\rho_{33}\geq|\rho_{23}|^2$ , $\rho_{11}\rho_{44}\geq|\rho_{14}|^2$, $\rho_{ii}\in \mathbb R$, $\rho_{23}= \rho_{32}^*$, and $\rho_{14}=\rho_{41}^*$.

Introduce real parameters $r=\rho_{11}-\rho_{44}+\rho_{22}-\rho_{33}$, $s=\rho_{11}-\rho_{44}-\rho_{22}+\rho_{33}$,
$c_3=\rho_{11}+\rho_{44}-\rho_{22}-\rho_{33}$, and complex variables $c_1=2(\rho_{23}+\rho_{14})$ and $c_2=2(\rho_{23}-\rho_{14})$.
Suppose the real and imaginary parts of $c_i$ are $a_i$ and $b_i$ ($i=1, 2$):
\begin{align*}
c_i=a_i+\sqrt{-1}b_i.
\end{align*}
Then the Bloch form of $\rho_{AB}$ is
\begin{align}\notag
&\rho_{AB}\\
&=\frac14[I+c_3\sigma_3\otimes\sigma_3+\sum\limits_{i=1,2} a_i\sigma_i\otimes\sigma_i+sI\otimes\sigma_3+r\sigma_3\otimes I+b_2\sigma_1\otimes\sigma_2-b_1\sigma_2\otimes\sigma_1],
\end{align}
where $\sigma_i$ are the Pauli spin matrices:
\begin{align*}
\sigma_1=\begin{pmatrix} 0 & 1\\ 1 & 0\end{pmatrix}, \sigma_2=\begin{pmatrix} 0 & -\sqrt{-1}\\ \sqrt{-1} & 0\end{pmatrix},  \sigma_3=\begin{pmatrix} 1 & 0\\ 0 & -1\end{pmatrix}
\end{align*}

It is easy to compute the eigenvalues of $\rho_{AB}$:
\begin{align}
\lambda_{1,2}=\frac{1}{4}(1+c_3\pm\sqrt{(r+s)^2+(a_1-a_2)^2+(b_1-b_2)^2}), \\
\lambda_{3,4}=\frac{1}{4}(1-c_3\pm\sqrt{(r-s)^2+(a_1+a_2)^2+(b_1+b_2)^2}).
\end{align}
The marginal state of $\rho_{AB}$ are then given by
\begin{align*}
\rho_A&=diag(\frac{1}{2}(1+r), \frac12 (1-r)),\\
\rho_B&=diag(\frac{1}{2}(1+s), \frac12 (1-s)).
\end{align*}

For $|y|\leqslant 1$, define the entropic function
\begin{align}\label{e:entropy}
E(y)=1-\frac12 (1+y)\log_2(1+y)-\frac12 (1-y)\log_2(1-y).
\end{align}
Here the values at $E(\pm 1)$ are taken as  $\lim_{y\to \pm 1}E(y)=0$.
Then the von Neumann entropies $S(\rho_A)$ and $S(\rho_B)$ are given by
\begin{align*}
S(\rho_A)&=E(r), \\  %\frac{1}{2}\sum_{\epsilon=\pm}(1+\epsilon r)\log_2(1+\epsilon r),\\
S(\rho_B)&=E(s).          %\frac{1}{2}\sum_{\epsilon=\pm}(1+\epsilon s)\log_2(1+\epsilon s).
\end{align*}

%\begin{align*}
%S(\rho_A)&=1-\frac{1}{2}\sum_{\epsilon=\pm}(1+\epsilon r)\log_2(1+\epsilon r),\\
%S(\rho_B)&=1-\frac{1}{2}\sum_{\epsilon=\pm}(1+\epsilon s)\log_2(1+\epsilon s).
%\end{align*}

With these quantities, the quantum mutual information is computed as
\begin{align}\notag%\label{Eq2.4}
\mathcal I(\rho_{AB})&=S(\rho_A)+S(\rho_B)+\sum_{i=1}^{4}\lambda_i\log_2\lambda_i\\ \notag
&=-4+E(r)+E(s)+E(c_3)\\ \label{e:qmutual}
&+\frac{1+c_3}2E(\frac{\sqrt{(r+s)^2+|c_1-c_2|^2}}{1+c_3})+\frac{1-c_3}2E(\frac{\sqrt{(r-s)^2+|c_1+c_2|^2}}{1-c_3}).
\end{align}

The weak measurements $\{P^B(x)\}$ are associated with  $\Pi_i$, which can be parameterized through the the special unitary group $\mathrm{SU}(2)$.
Up to a phase factor, any element $V$ of $\mathrm{SU}(2)$ can be written as
$V=tI+i\sum_{i=1}^3y_i\sigma_i$, where $t, y_1, y_2, y_3$ are real numbers such that $t^2+y_1^2+y_2^2+y_3^2=1$.
One can directly compute that
\begin{equation}\label{Eq2.7}
V^\dag\sigma_1V=(t^2+y_1^2-y_2^2-y_3^2)\sigma_1+2(ty_3+y_1y_2)\sigma_2+2(-ty_2+y_1y_3)\sigma_3, \\
\end{equation}
and $V^\dag\sigma_2V$, $V^\dag\sigma_3V$ are obtained from \eqref{Eq2.7}
under the cyclic permutations $(\sigma_1, \sigma_2, \sigma_3)\mapsto (\sigma_2, \sigma_3, \sigma_1)$
and $(y_1, y_2, y_3)\mapsto (y_2, y_3, y_1)$.

Let
$$z_1=2(-ty_2+y_1y_3), \quad z_2=2(ty_1+y_2y_3), \quad z=t^2+y_3^2-y_1^2-y_2^2. $$
Then $z_1^2+z_2^2+z^2=1$, thus $|z|\leqslant 1$.
It follows from (\ref{c}) that
\begin{align}\label{e:rhoA}
&\rho_{A|P^B(x)}\\ \notag
&=\frac{I(1-sz\tanh x)+(r-c_3z\tanh x)\sigma_3-[(z_1a_1+z_2b_2)\sigma_1+(z_2a_2-z_1b_1)\sigma_2]\tanh x}{2(1-sz\tanh x)}
%\\
%&\rho_{A|P^B(-x)}\notag\\
%&=\frac{I(1+sz\tanh x)+(r+c_3z\tanh x)\sigma_3+[(z_1a_1+z_2b_2)\sigma_1+(z_2a_2-z_1b_1)\sigma_2]\tanh x}{2(1+sz\tanh x)}
\end{align}
and $\rho_{A|P^B(-x)}$ is given by replacing $x$ with $-x$ in \eqref{e:rhoA}. Here
$p(\pm x)=\frac{1}{2}(1\mp sz\tanh x)$. %, p(-x)=\frac{1}{2}(1+sz\tanh x)$.

The eigenvalues of $\rho_{A|P^B(x)}$ and $\rho_{A|P^B(-x)}$ are given by£º
\begin{align}\label{e:rho}
\lambda^{\pm}_{\rho_{A|P^B(x)}}&=\frac{1-sz\tanh x\pm\sqrt{r^2-2rzc_3\tanh x+\theta\tanh^2x}}{2(1-sz\tanh x)}\\ \label{e:rho2}
%\lambda^{\pm}_{\rho_{A|P^B(+x)}}&=\frac{1-sz\tanh x\pm\sqrt{r^2-2rzc_3\tanh x+a^2\tanh^2x+c_3^2z^2\tanh^2x}}{2(1-sz\tanh x)},\\
\lambda^{\pm}_{\rho_{A|P^B(-x)}}&=\frac{1+sz\tanh x\pm\sqrt{r^2+2rzc_3\tanh x+\theta\tanh^2x}}{2(1+sz\tanh x)},
\end{align}
where we have introduced a new variable $\theta$ by
\begin{equation}
\theta=z_1^2|c_1|^2+z_2^2|c_2|^2+2z_1z_2|c_1\times c_2|+c_3^2z^2.
\end{equation}
Recall that $c_1, c_2$ are given complex numbers,
and $c_1\times c_2=(a_1b_2-a_2b_1)\overrightarrow{k}$.

%$a^2=z_1^2|c_1|^2+z_2^2|c_2|^2+2z_1z_2|c_1\times c_2|$. Note that $
%To find the minimum value in the super quantum discord, we first consolidate the variables to define

Note that formulas \eqref{e:rho}-\eqref{e:rho2} always give real numbers as $\theta\geqslant c_3^2z^2$, and they also imply that
\begin{align}\label{e:region}
r^2\pm 2rzc_3\tanh x+\theta\tanh^2x\leqslant (1\pm sz\tanh x)^2,
\end{align}
which shows that $\theta$ is bounded above.
%In particular, $\sqrt{r^2+b^2\tanh^2x}\leqslant 1$, where $b^2=\frac{|c_1|^2+|c_2|^2+\sqrt{(|c_1|^2-|c_2|^2)^2+4|c_1\times c_2|^2}}{2}$.

To calculate the super quantum discord of $\rho_{AB}$ using (\ref{c}) and  (\ref{superd}), we need to calculate the classical correlation and minimize $S(A|\{P^B(x)\})$ with respect to the weak measurements
$\{P^B(\pm x)\}$. Using the formulas for the eigenvalues we find out that
\begin{equation}
\min_{\{P^B(x)\}}S(A|\{P^B(x)\})=1+\min_{z,\theta}G(\theta,z)\\
\end{equation}
with
\begin{align*}\label{g}
G(\theta, z)=&-\frac{1}{4}(1+sz\tanh x+R_+)\log_2\frac{1+sz\tanh x+R_+}{1+sz\tanh x}\\
&-\frac{1}{4}(1+sz\tanh x-R_+)\log_2\frac{1+sz\tanh x-R_+}{1+sz\tanh x}\\
&-\frac{1}{4}(1-sz\tanh x+R_-)\log_2\frac{1-sz\tanh x+R_-}{1-sz\tanh x}\\
&-\frac{1}{4}(1-sz\tanh x-R_-)\log_2\frac{1-sz\tanh x-R_-}{1-sz\tanh x},\\
\end{align*}
where $R_\pm=\sqrt{r^2\pm2rc_3z\tanh x+\theta\tanh^2x}$. The minimum is taken over a 2-dimensional region such that $|z|\leqslant 1$ and $\theta$ is implicitly bounded by
\eqref{e:region}.

Observe that $G(\theta, -z)=G(\theta, z)$, so we only need to consider $z\in[0,1]$. Furthermore, we can reduce the optimization of the two variable function
to that of one variable.

\begin{lemma}\label{le:1} Let $b^2=\frac{|c_1|^2+|c_2|^2+\sqrt{(|c_1|^2-|c_2|^2)^2+4|c_1\times c_2|^2}}{2}$, then the minimum of the quantity $S(A|\{P^B(x)\})$ is given by
\begin{equation}\label{e:min}
\min_{\{P^B(x)\}}S(A|\{P^B(x)\})=1+\min_{z\in[0,1]}F(z),
\end{equation}
where
\begin{equation}\label{e:F}
\begin{aligned}\notag
F(z)=&-\frac{1}{4}(1+sz\tanh x+H_+)\log_2\frac{1+sz\tanh x+H_+}{1+sz\tanh x}\\
&-\frac{1}{4}(1+sz\tanh x-H_+)\log_2\frac{1+sz\tanh x-H_+}{1+sz\tanh x}\\
&-\frac{1}{4}(1-sz\tanh x+H_-)\log_2\frac{1-sz\tanh x+H_-}{1-sz\tanh x}\\
&-\frac{1}{4}(1-sz\tanh x-H_-)\log_2\frac{1-sz\tanh x-H_-}{1-sz\tanh x}
\end{aligned}
\end{equation}
and $H_{\pm}=\sqrt{b^2(1-z^2)\tanh^2x+(r\pm c_3z\tanh x)^2}$.%=\sqrt{(r^2+c^2\tanh^2x)\pm 2rc_3z\tanh x+(c_3^2-c^2)z^2}\tanh^2x$.
\end{lemma}
See Appendix for a proof.

%This lemma gives a new method to reduce
%the associated optimization problem into that of a {\it one-variable} entropy-like function
%on the closed interval $[0, 1]$, which in principle solves the problem of the $\min_{\{P^B(x)\}}S(A|\{P^B(x)\})$'s value.
%This theorem completes settles the problem of super quantum discord by well-known result of calculus. We have the following analytic formulas for several
%regions of the seven parameters.

\begin{theorem}\label{t:1} The super quantum discord of the
general two qubit $X$-state $\rho_{AB}$ is
\begin{align}\notag
SD(\rho_{AB})=&S(\rho_B)-S(\rho_{AB})+\min_{\{P^B(x)\}}S(A|P^B(x))\\ \notag
=&E(s)+E(c_3)+\frac{1+c_3}2E(\frac{\sqrt{(r+s)^2+|c_1-c_2|^2}}{1+c_3})\\ \label{sqd}
&+\frac{1-c_3}2E(\frac{\sqrt{(r-s)^2+|c_1+c_2|^2}}{1-c_3})-3+\min_{z\in[0,1]}F(z)
\end{align}
where %$S(\rho_{B}), S(\rho_{AB})$ are the von Neumann entropies of the reduced state $\rho_B$ and the total state $\rho_{AB}$,
$E(y)$ is the entropic function defined in \eqref{e:entropy} and $F(z)$ is given by  Lemma \ref{le:1}.
\end{theorem}

The above result essentially determines the quantum super discord completely, as it is expressed as the minimum of one-variable function
$F(z)$ on $[0, 1]$. For a given $x$, the function $F(z)$ depends on the complex parameters $c_1, c_2$ and there real parameters $r, s, c_3$, therefore $F(z)$ lives
on a 7-dimensional manifold. Theorem \ref{t:1} effectively reduces the parameters to 4 real ones $b, c_3, r, s$ together with the measurement
parameter $x$. In the following we give exact results for several nontrivial regions of the parameters of the quantum state $\rho_{AB}$.

\begin{corollary}\label{C:1} For the general two qubit $X$-type quantum state, the super quantum discord is explicitly computed according to the following cases.

(a) If  $s\tanh x \geqslant0$, $rc_3\tanh x \leqslant0$ and $c_3^2-b^2\geqslant src_3$, %or $s\tanh x=0, c_3^2\geqslant b^2$,
then the super quantum discord is given by  (\ref{sqd}) with $\min_{z\in[0, 1]} F(z)=F(1)$, and

\begin{equation}\label{e:f1}
\begin{aligned}
F(1)=&-\frac{1}{4}(1+r+(s+c_3)\tanh x)\log_2\frac{1+r+(s+c_3)\tanh x}{1+s\tanh x}\\
&-\frac{1}{4}(1-r+(s-c_3)\tanh x)\log_2\frac{1-r+(s-c_3)\tanh x}{1+s\tanh x}\\
&-\frac{1}{4}(1+r-(s+c_3)\tanh x)\log_2\frac{1+r-(s+c_3)\tanh x}{1-s\tanh x}\\
&-\frac{1}{4}(1-r-(s-c_3)\tanh x)\log_2\frac{1-r-(s-c_3)\tanh x}{1-s\tanh x}
\end{aligned}
\end{equation}

(b) If $s\tanh x\leqslant0$, $rc_3\tanh x\geqslant0$ and $c_3^2-b^2\geqslant src_3$, then the super quantum discord is given by the same formula as in (a).

(c) If $r=s=0$ and $c_3^2\leq b^2$, then the super quantum discord is given by  (\ref{sqd}) with $\min_{z\in[0, 1]} F(z)=F(0)$, where
\begin{align}\label{case-d}
\begin{aligned}
F(0)=&E(b\tanh x)-1\\
=&-\frac{1}{2}(1+b\tanh x)\log_2(1+b\tanh x)\\
&-\frac{1}{2}(1-b\tanh x)\log_2(1-b\tanh x).
\end{aligned}
\end{align}

(d) If $s=rc_3$, $b^2=c_3^2$, and $r^2+c_3^2\tanh^2x\pm rc_3\tanh x\geq1$, then the super quantum discord is given by  (\ref{sqd}) where $\min_{z\in[0, 1]} F(z)=F(0)$, where
\begin{align}
\begin{aligned}
F(0)=&E(\sqrt{r^2+c_3^2\tanh^2 x})-1\\
=&-\frac{1}{2}(1+\sqrt{r^2+c_3^2\tanh^2 x})\log_2(1+\sqrt{r^2+c_3^2\tanh^2 x})\\
&-\frac{1}{2}(1-\sqrt{r^2+c_3^2\tanh^2 x})\log_2(1-\sqrt{r^2+c_3^2\tanh^2 x}).
\end{aligned}
\end{align}

\end{corollary}
See Appendix for a proof.

Remark. The above corollary shows that the super quantum discord is mostly determined by $F(1)$, but there are still other solutions not covered
by this result. For example, Example 3 below is not covered by the above result, and cannot be solved by any existing
algorithms.

It is imperative to find a new method to
resolve the situation.  The following formula will fill up the gaps
in the literature, and covers all the situations for the general X-state.

\begin{theorem} The exceptional optimal points of $F(z)$ are determined by the iterative formula:
\begin{equation}\label{newton}
\hat{z}=\lim_{n\to\infty}(z_n-\frac{F'(z_n)}{F''(z_n)}),
\end{equation}
where
\begin{align}\label{e:difb1}
F'(z)=-\frac{1}{4\ln2}&\left\{s\tanh x\ln\frac{((1+sz\tanh x)^2-H_+^2)(1-sz\tanh x)^2}{((1-sz\tanh x)^2-H_-^2)(1+sz\tanh x)^2}\right.\notag\\
+H_+^{\prime}&\left.\ln\frac{1+sz\tanh x+H_+}{1+sz\tanh x-H_+}+H_-^{\prime}\ln\frac{1-sz\tanh x+H_-}{1-sz\tanh x-H_-}\right\}\\
F''(z)=-\frac1{2\ln 2}&\left\{\frac{(s^2\tanh^2x+H_+'^2)(1+sz\tanh x)-2s\tanh xH_+H_+'}{(1+sz\tanh x)^2-H_+^2}\right.\notag\\
&\left.+\frac{(s^2\tanh^2x+H_-'^2)(1-sz\tanh x)+2s\tanh xH_-H_-'}{(1-sz\tanh x)^2-H_-^2}\right.\notag\\
&-\left.\frac{2s^2\tanh^2x}{1-s^2z^2\tanh^2x}+\frac12 H_+''\ln\frac{1+sz\tanh x+H_+}{1+sz\tanh x-H_+}+\frac12 H_-''\ln\frac{1-sz\tanh x+H_-}{1-sz\tanh x-H_-}\right\}.
\end{align}

As $F'(0)=F^{(3)}(0)$, the iteration usually starts with $z_0=1$.
\end{theorem}

\medskip

Example 1. Let $\rho=\frac14(I+\sum_{i=1}^3 c_i\sigma_i\otimes \sigma_i)$ be the Bell-diagonal state. Then $r=s=0$. This is a special
case of Corollary \ref{C:1}, so the minimum of $F(z)$ on $[0,1]$ is $F(0)$ or $F(1)$. If $c_3^2\geq b^2$, then $\min_{z\in[0, 1]} F(z)=F(1)=-\frac{1}{2}(1+c_3\tanh x)\log_2(1+c_3\tanh x)-\frac{1}{2}(1-c_3\tanh x)\log_2(1-c_3\tanh x)$;
If $c_3^2\leq b^2$, then $\min_{z\in[0, 1]} F(z)=F(0)=-\frac{1}{2}(1+b\tanh x)\log_2(1+b\tanh x)-\frac{1}{2}(1-b\tanh x)\log_2(1-b\tanh x)$. Thus the super quantum discord of $\rho$ is
\begin{equation}
\begin{split}
SD(\rho)&=\frac{1}{4}(1-c_3+c_1+c_2)\log_2(1-c_3+c_1+c_2)\\
&+\frac{1}{4}(1-c_3-c_1-c_2)\log_2((1-c_3-c_1-c_2)\\
&+\frac{1}{4}(1+c_3+c_1-c_2)\log_2((1+c_3+c_1-c_2)\\
&+\frac{1}{4}(1+c_3-c_1+c_2)\log_2((1+c_3-c_1+c_2)\\
&-\frac{1}{2}(1+C\tanh x)\log_2(1+C\tanh x)-\frac{1}{2}(1-C\tanh x)\log_2(1-C\tanh x).
\end{split}
\end{equation}
This solution was first given in \cite{W} with $C=\max\{|c_3|,|b|\}$.

Note that the Werner state $
\rho=a|\psi^-\rangle\langle\psi^-|+\frac{1-a}{4}I$,
$0\leq a\leq1$, is a special case with $r=s=0,c_3=-a,c_1=c_2=-a$. Here $|\psi^-\rangle=(|01\rangle-|10\rangle)/\sqrt{2}$.

\medskip

Example 2. Let $\rho$ be the following density matrix:
\begin{equation}\label{f1}
\begin{pmatrix}
       0.437&0&0&0.100\\
       0&0.154&0&0\\
       0&0&0.037&0\\
       0.100&0&0&0.372\\
\end{pmatrix}.
\end{equation}
Here $r=0.182, s=-0.052, c_3=0.618, c_1=0.2, c_2=-0.2$, so $b=0.2$. One sees that this belongs to Corollary \ref{C:1} (a) and (b), so $\min_{z\in[0, 1]} F(z)=F(1)$. Fig.1 shows that $F(z)$ as a function of $x\geq0$ and $z\in[0,1]$, we can observe the behaviour of $F(z)$ more intuitively. The eigenvalues of $\rho$ are $\lambda_1=0.509649,\lambda_2=0.299351,\lambda_3=0.154,\lambda_4=0.037$. Following  (\ref{sqd}), the super quantum discord of $\rho$ is given by
\begin{equation}
\begin{aligned}
SD(\rho)=&2-\frac{1}{2}(1+s)\log_2(1+s)-\frac{1}{2}(1-s)\log_2(1-s)\\
&+\sum_{i=1}^{4}\lambda_i\log_2\lambda_i
+F(1)=0.3899+F(1).
\end{aligned}
\end{equation}
where
\begin{equation}
\begin{aligned}\notag
F(1)=&-\frac{1}{4}(1.182+0.566\tanh x)\log_2\frac{1.182+0.566\tanh x}{1-0.052\tanh x}\\
&-\frac{1}{4}(0.818-0.64\tanh x)\log_2\frac{0.818-0.64\tanh x}{1-0.052\tanh x}\\
&-\frac{1}{4}(1.182-0.566\tanh x)\log_2\frac{1.182-0.566\tanh x}{1+0.052\tanh x}\\
&-\frac{1}{4}(0.818+0.64\tanh x)\log_2\frac{0.818-0.64\tanh x}{1+0.052\tanh x}
\end{aligned}
\end{equation}

\begin{figure}[!h]
  \centering
  \includegraphics[width=3.5in]{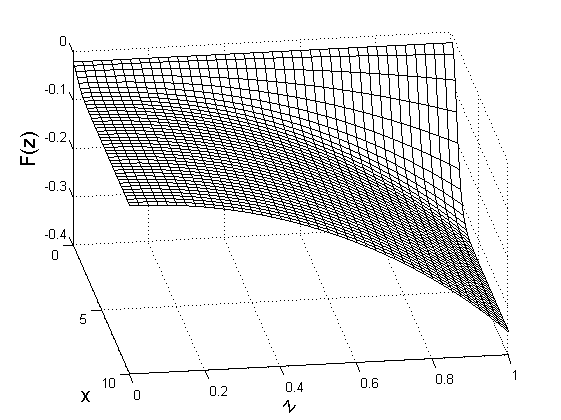}\\
  \caption{The behaviour of $F(z)$ for $x\geq0$ and $z\in[0,1]$ with parameters $r=0.182, s=-0.052, c_3=0.618, b=0.2$.}
\end{figure}

One can easily prove that $F(1)$ is an even function of $x$, and when $x\geq0$, $F(1)$ decreases with increasing $x$. Which means the super quantum discord of
this state $\rho$ is a monotonically decreasing function of the measurement strength. This is consistent with the Theorem.2 in \cite{SP}.

\medskip

Example 3. The following example cannot be solved by any of the currently available algorithms until this paper. Using our new method, its exact solution is obtained as follows.  Consider the density matrix $\rho$ given by
\begin{equation}\label{counterex}
\begin{pmatrix}
       0.0783&0&0&0\\
       0&0.1250&0.1000&0\\
       0&0.1000&0.1250&0\\
       0&0&0&0.6717\\
\end{pmatrix}.
\end{equation}
In terms of the Bloch form, $r=s=-0.5934,c_3=0.5,c_1=c_2=0.2$, so $b=0.2$. The eigenvalues of $\rho$ are $\lambda_1=0.025,\lambda_2=0.0783,\lambda_3=0.2250,\lambda_4=0.6717$. By symmetry, it is enough to consider $x>0$.
The function $F(z)$ in deciding the super quantum discord is shown on the left side of Fig. 1 as a function of $x\geq0$ and $z\in[0,1]$. The right side of Fig. 2 shows contour pictures of $F(z)$ by choosing $x=1,2,3,4$. The red dot on each line is the minimum of $F(z)$. The graphs reveal that the optimal point $\hat{z}\neq 0, 1$, which
can also be computed explicitly by  (\ref{newton}). This example shows that the claim that the maximum is always given at either $z=0$ or
$z=1$ is incorrect (cf. \cite{LT}), see the third and fourth graphs in Fig. 2 for more information.

For example, set $x=3$. Starting with with $z_0=1$, (\ref{newton}) gives that $z_1=0.8305, z_2=0.6718, z_3=0.5582, z_4=0.4964, z_5=0.4788, z_6=0.477467, z_7=0.4774675, z_8=0.4774676\ldots $,
thus $\hat{z}=0.47747$ is the optimal point of $F(z)$. It follows from (\ref{sqd}) that the super quantum discord of $\rho$ is $SD(\rho)=2-\frac{1}{2}(1+s)\log_2(1+s)-\frac{1}{2}(1-s)\log_2(1-s)+\sum_{i=1}^{4}\lambda_i\log_2\lambda_i+
F(0.47747)=0.1332$.

Similarly if $x=4$, $z_0=1$,
 (\ref{newton}) gives that $z_1=0.9042, z_2=0.8561, z_3=0.8467, z_4=0.84638901, z_5=0.846388659...$,
thus $\hat{z}=0.84639$ is another critical point of $F(z)$.
Finally the super quantum discord turns out to be
$SD(\rho)=2-\frac{1}{2}(1+s)\log_2(1+s)-\frac{1}{2}(1-s)\log_2(1-s)+\sum_{i=1}^{4}\lambda_i\log_2\lambda_i+
F(0.84639)=0.1328$.

\begin{figure}[!h]
  \centering
  \includegraphics[width=6.5in]{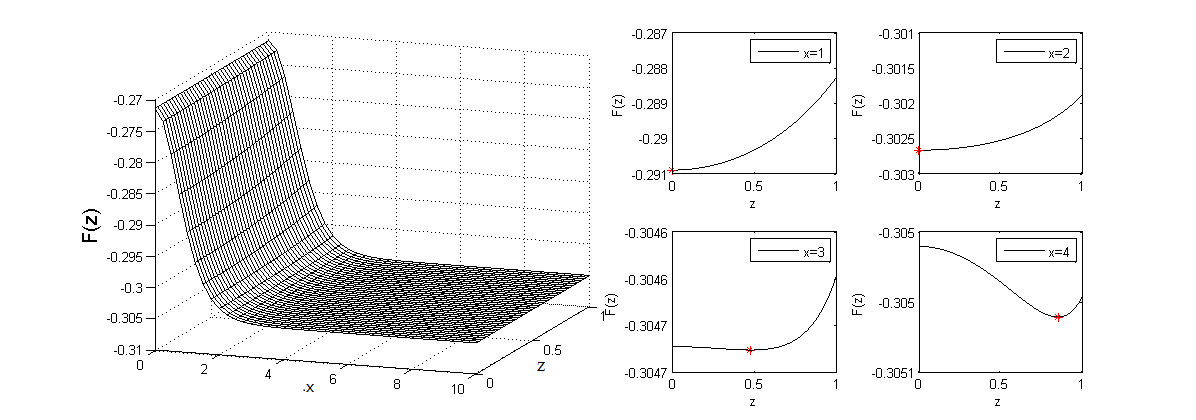}\\
  \caption{The behaviour of $F(z)$ for $x\geq0$ and $z\in[0,1]$ with parameters $r=-0.5934, s=-0.5934, c_3=0.5, b=0.2$. The red dot on each line represents the optimal point. The 3rd and 4th graphs show that the maximum is not given by $z=0$ or $z=1$, instead they are
  respectively given by $\hat z=0.47747$ and $\hat z=0.84639$ as shown by the red dots.}\label{1}
\end{figure}

\medskip

\section{DYNAMICS of SUPER QUANTUM DISCORD under PHASE DAMPING CHANNEL}

In this section, we discuss the behavior of the general 2-qubit $X$ state $\rho_{AB}$ through the phase damping channels \cite{NC} with the Kraus operators $\{K_i\}$, where $\sum_iK_i^\dag K_i=1$. Under the phase damping $\rho_{AB}$ evolves into
\begin{equation}
\tilde{\rho}_{AB}=\sum_{i,j\in{{1,2}}}K_i^A\otimes K_j^B\cdot\rho_{AB}\cdot (K_i^A\otimes K_j^B)^\dag.
\end{equation}
where the Kraus operators are given by
 $K_1^{A(B)}=\ket{0}\bra{0}+\sqrt{1-\gamma}\ket{1}\bra{1}$, and $K_2^{A(B)}=\sqrt{\gamma}\ket{1}\bra{1}$, with the decoherence rate $\gamma\in[0,1]$. Thus we have
\begin{align}\label{rho}
\tilde{\rho}=\frac14[&I+c_3\sigma_3\otimes\sigma_3+sI\otimes\sigma_3+r\sigma_3\otimes I
+\sum\limits_{i=1,2}a_i(1-\gamma)\sigma_i\otimes\sigma_i\\ \notag
+&b_2(1-\gamma)\sigma_1\otimes\sigma_2-b_1(1-\gamma)\sigma_2\otimes\sigma_1],
\end{align}

The parameter $\gamma$ also determines how severely the noise in the channel affects the super quantum discord. Clearly, when $\gamma=0$, super quantum discord is preserved.  As we have mentioned that
the super quantum discord tends to decrease when the strength $x$ increases \cite[Th. 2]{SP}. It is interesting to see
the behavior of $SD(\tilde{\rho})$ under the phase damping channel, for example, how does the SQD of the X-state change for general $\gamma$?
\medskip

(1) First we consider the Werner state $\rho=a|\psi^-\rangle\langle\psi^-|+\frac{1-a}{4}I$, $a\in[0, 1]$.
Under the phase damping channel, the Werner state $\rho$ is changed into
\begin{equation}
\tilde{\rho}
=\frac14(I-a\sigma_3\otimes\sigma_3-a(1-\gamma)\sum_{i=1, 2}\sigma_i\otimes\sigma_i)
\end{equation}
Then $r=s=0, c_3=-a, c_1=c_2=-a(1-\gamma)$, $b=|a(1-\gamma)|$ (see Lemma \ref{le:1}). It follows from $c_3^2\geq b^2$ that the super quantum discord of $\tilde{\rho}$ is
\begin{equation}
\begin{split}
SD(\tilde{\rho})&=\frac{1}{4}(1+a-2a(1-\gamma))\log_2(1+a-2a(1-\gamma))\\
&+\frac{1}{4}(1+a+2a(1-\gamma))\log_2(1+a+2a(1-\gamma))+\frac{1}{2}(1-a)\log_2(1-a)\\
&-\frac{1}{2}(1-a\tanh x)\log_2(1-a\tanh x)-\frac{1}{2}(1+a\tanh x)\log_2(1+a\tanh x).
\end{split}
\end{equation}
%\begin{equation}
%\begin{split}
%SD(\rho)&=\frac{1}{4}(1-a)\log_2(1-a)\\
%&+\frac{1}{4}(1+3a)\log_2(1+3a)\\
%&+\frac{1}{2}(1-a)\log_2(1-a)\\
%&-\frac{1}{2}(1-a\tanh x)\log_2(1-a\tanh x)-\frac{1}{2}(1+a\tanh x)\log_2(1+a\tanh x).
%\end{split}
%\end{equation}
Thus
\begin{equation}
\begin{split}
SD(\rho)-SD(\tilde{\rho})&=\frac{1}{4}(1-a)\log_2(1-a)+\frac{1}{4}(1+3a)\log_2(1+3a)\\
&-\frac{1}{4}(1+a-2a(1-\gamma))\log_2(1+a-2a(1-\gamma))\\
&-\frac{1}{4}(1+a+2a(1-\gamma))\log_2(1+a+2a(1-\gamma)).
\end{split}
\end{equation}

Set $T(a,\gamma)=SD(\rho)-SD(\tilde{\rho})$, which is a strictly increasing function of $\gamma$:
%Fig.3 shows that $T(a,\mu)$ as a function of $a\in[0,1]$ and $\mu\in[0,1]$.
%\begin{figure}[!h]
 % \centering
  %\includegraphics[width=3.5in]{fig4}\\
  %\includegraphics{li}\\
  %\caption{The behaviour of $T(a,\mu)$ for $a\in[0,1]$ and $\mu\in[0,1]$.}
%\end{figure}
%Because of $SD(\rho)\geq SD(\tilde{\rho})$, we can conclude that the super quantum discord of Werner state decreases under phase damping channel.
\begin{equation}
\frac{\partial T(a,\gamma)}{\partial\gamma}=\frac{a}{2}\log_2(\frac{1+a+2a(1-\gamma)}{1+a-2a(1-\gamma)})\geq0.
\end{equation}
Thus for fixed $a\in [0, 1]$, the minimum of $T(a,\gamma)$ takes place at $\gamma=0$. Since $T(a,0)=0$, we conclude that $\min T(a,\gamma)=0$. Therefore
$SD(\rho)\geq SD(\tilde{\rho})$,  which implies that the super quantum discord of the Werner state decreases under the phase damping channel.

(2) As another example, let us consider the state $\rho$ discussed in Example 2. The state $\rho$ under phase damping channel is given by
\begin{equation}
\tilde{\rho}=\frac14[I+0.618\sigma_3\otimes\sigma_3-0.052I\otimes\sigma_3+0.182\sigma_3\otimes I
+0.2(1-\gamma)\sigma_1\otimes\sigma_1-0.2(1-\gamma)\sigma_2\otimes\sigma_2].
\end{equation}

 Therefore, its Bloch form is given by $r=0.182, s=-0.052, c_3=0.618, c_1=0.2(1-\gamma), c_2=-0.2(1-\gamma)$, $b=\max\{|c_1|,|c_2|\}=0.2(1-\gamma)$. After a simple calculation, we obtain that the parameters in this case satisfy the conditions of Corollary \ref{C:1} $(a)$ and $(b)$, thus the minimum of $F(z)$ is also $F(1)$. From Theorem \ref{t:1}, we obtain that
$$SD(\tilde{\rho})=SD(\rho), $$
which says that there are essentially no noises detected for this particular quantum state in the phase damping channel for the super quantum discord in the final stage.
\medskip

(3) We consider the state $\rho$ discussed in Example 3. From  (\ref{rho}), the state $\rho$ under phase damping channel evolves into
\begin{equation}
\tilde{\rho}=\frac14[I+0.5\sigma_3\otimes\sigma_3-0.5934I\otimes\sigma_3-0.5934\sigma_3\otimes I
+\sum\limits_{i=1,2}0.2(1-\gamma)\sigma_i\otimes\sigma_i].\\
\end{equation}
Here $r=s=-0.5934, c_3=0.5, c_1=c_2=0.2(1-\gamma)$, so $b=0.2(1-\gamma)$.
From Example 3, we find that when $x=4$, the optimal point $\hat{z}\in (0, 1)$, and that is, the minimum of $F(z)$ is $F(\hat{z})=F(0.8464)$. The SQD of the state $\rho$ is exactly calculated in Ex.3.

Fig. 3 below illustrates $F(z)$ as a function of $z\in[0,1]$ and $\gamma\in[0,1]$ with fixed parameters $r=-0.5934, s=-0.5934, c_3=0.5, x=4$.  The solid lines in the last graph from top to bottom correspond to $\gamma=1,0.7,0.5,0.3,0.1$ respectively. We find out that when parameter $\gamma\in(0,1]$, the optimal point is $z=1$, and $F(1)>F(\hat z)$.
From Theorem \ref{t:1}, we obtain $SD(\tilde{\rho})\geq SD(\rho)$, which means in this case the super quantum discord of the state $\rho$ is destroyed through the phase damping channel.
\begin{figure}[!h]
  \centering
  \includegraphics[width=5.5in]{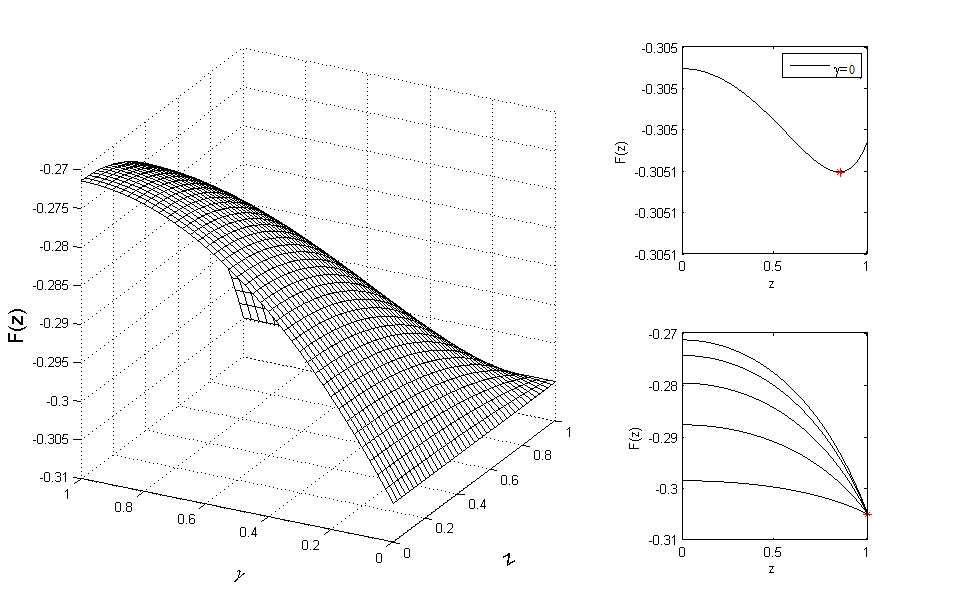}\\
  \caption{The solid lines in the last two graphs from top to bottom correspond to $\gamma=1,0.7,0.5,0.3,0.1$ respectively, for fixed parameters $r=-0.5934, s=-0.5934, c_3=0.5, x=4$. The red dot is the optimal point. Again the optimal is achieved not at endpoints.}
\end{figure}

These models show that usually the super quantum discord decreases as the decoherence rate increases. However, there also exists an example
when the super quantum discord is stable through the whole process under damping channel.

\section{Conclusions}

In this paper we have derived an analytical solution for the super quantum discord of the general two qubit $X$-states in terms of the minimum of certain one variable
function $F(z)$ on $[0, 1]$. Although usually  the super quantum discord is given by either $F(0)$ or $F(1)$ (see Corollary \ref{C:1}), there are
significantly many other cases. % reveals that, which agrees with most of the results given by \cite{LT}.
We have formulated an effective algorithm to nail down theses exceptional cases, which have
filled up the gap in the literature for super quantum discords.
%For those optimal points that have been missed in the leterature,
%we have given an iterative formula to compute them exactly.
Therefore our results have settled the problem of the super quantum discord
for the general two qubit $X$ states. In general the super quantum discord tends to increase or decrease under the phase damping channel, though there exists some extreme case when the super quantum discord is invariant through the whole dynamic process.
%We have analyzed how super quantum discord behaves under phase damping channel. Moreover,
%we have derived several interesting properties of the dynamic behavior of the state through phase damping channel.

\bigskip

\centerline{Acknowledgments}
The research is partially supported by
NSFC grant (Nos. 11271138, 11531004) and Simons Foundation (No. 198129).

\bigskip

\begin{center}
Appendix
\end{center}

Proof of Lemma 3.1.

First we notice that $G(\theta, z)$ is a strictly decreasing function of $\theta$:
\begin{align*}
\frac{\partial G}{\partial\theta}=&-\frac{\tanh^2x}{8R_+}\log_2(\frac{1+sz\tanh x+\sqrt{r^2+2rc_3z\tanh x+\theta\tanh^2 x}}{1+sz\tanh x-\sqrt{r^2+2rc_3z\tanh x+\theta\tanh^2x}})\\
&-\frac{\tanh^2x}{8R_-}\log_2(\frac{1-sz\tanh x+\sqrt{r^2-2rc_3z\tanh x+\theta\tanh^2x}}{1-sz\tanh x-\sqrt{r^2-2rc_3z\tanh x+\theta\tanh^2x}})\\
<&0.
\end{align*}
Therefore there are no interior critical points and extremal points must lie on the boundary
of the domain. Since $\frac{\partial G}{\partial\theta}<0$, we further conclude that
$\min G$ takes place at the largest value of $\theta$ for some $z\in [0, 1]$.
As $z_1^2+z_2^2+z^2=1$, we have that
\begin{align*}
\theta&=a^2+c_3^2z^2\\
&=z_1^2|c_1|^2+z_2^2|c_2|^2+2z_1z_2|c_1\times c_2|+c_3^2z^2\\
&\leqslant (\frac{|c_1|^2+|c_2|^2+\sqrt{(|c_1|^2-|c_2|^2)^2+4|c_1\times c_2|^2}}{2})(z_1^2+z_2^2)+c_3^2z^2\\
&=b^2+(c_3^2-b^2)z^2,
\end{align*}
where $b^2=\frac{|c_1|^2+|c_2|^2+\sqrt{(|c_1|^2-|c_2|^2)^2+4|c_1\times c_2|^2}}{2}$

For each fixed $z$, the maximum value $b^2+(c_3^2-b^2)z^2$ can be achieved by appropriate $z_1, z_2$.
Therefore $\min G(z, \theta)=\min_{z\in[0,1]}G(z, b^2+(c_3^2-b^2)z^2)$, which is
explicitly given in  (\ref{e:min}).

\bigskip

Proof of Theorem 3.2.

We compute the derivative of $F(z)$.
\begin{equation}\label{e:difb}
\begin{aligned}
F^{\prime}(z)=&-\frac{1}{4}\left(s\tanh x\log_2\frac{1-A_+^2}{1-A_-^2}
+\frac{rc_3\tanh x+(c_3^2-b^2)z\tanh^2x}{1+sz\tanh x}\frac1{A_+}\log_2\frac{1+A_+}{1-A_+}\right.\\
&\qquad\qquad+\left.\frac{-rc_3\tanh x+(c_3^2-b^2)z\tanh^2x}{1-sz\tanh x}\frac1{A_-}\log_2\frac{1+A_-}{1-A_-}\right)
\end{aligned}
\end{equation}
where $H_{\pm}^{\prime}(z)=H_\pm^{-1}(\pm rc_3\tanh x+(c_3^2-b^2)z\tanh^2x)$ and $\displaystyle A_{\pm}=\frac{H_\pm}{1\pm sz\tanh x}\in [0, 1]$.

{\it Case (a)}. Since
\begin{equation}\label{e:AA}
\begin{aligned}
&A_+^2-A_-^2=\frac{H_+^2}{(1+sz\tanh x)^2}-\frac{H_-^2}{(1-sz\tanh x)^2}\\
%=&\frac{(1-s\tanh xz)^2((r+c_3\tanh xz)^2+b^2(1-z^2)\tanh^2x)}{(1-s\tanh xz)^2(1+s\tanh xz)^2}\\
%&-\frac{(1+s\tanh xz)^2((r-c_3\tanh xz)^2+b^2(1-z^2)\tanh^2x)}{(1-s\tanh xz)^2(1+s\tanh xz)^2}\\
=&\frac{(1+s^2z^2\tanh^2x)4rc_3z\tanh x-4sz\tanh x[(r^2+c_3^2z^2\tanh^2x)+b^2\tanh^2x(1-z^2)]}{(1-sz\tanh x)^2(1+sz\tanh x)^2}
%&=\frac{4sz\left((c^2-c_3^2+src_3)z^2+s^{-1}rc_3-r^2-c^2\right)}{(1-sz)^2(1+sz)^2}.
\end{aligned}
\end{equation}

Then the first term of $F'(z)\leq 0$ iff $-s\tanh x(A_+^2-A_-^2)\geqslant 0$, which holds if
$s\tanh x\geqslant 0$ and $rc_3\tanh x\leqslant 0$ or $s\tanh x\leqslant 0$ and $rc_3\tanh x\geqslant 0$. %In particular, $r=0$
%implies that $-s\tanh x(A_+^2-A_-^2)\geqslant 0$.

Note that $\displaystyle g(x)=\frac1x\ln\frac{1+x}{1-x}$ is a strictly increasing function on $(0, 1)$, as
\begin{align*}
g'(x)&=-\frac1{x^2}\log_2\frac{1+x}{1-x}+\frac2{x\ln 2}\frac1{1-x^2}\\
&=\frac{2}{x\ln 2}\left(\sum_{n=0}^{\infty}\frac{-x^{2n}}{2n+1}+\sum_{n=0}^{\infty}x^{2n}\right)>0.
\end{align*}
Therefore
$A_+\geqslant A_-$ iff
\begin{equation}\label{e:logA}
\frac1{A_+}\ln\frac{1+A_+}{1-A_+}\geqslant \frac1{A_-}\ln\frac{1+A_-}{1-A_-}.
\end{equation}

(i) If $s\tanh x\geqslant 0,rc_3\tanh x\leqslant 0$ and $c_3^2-b^2\geqslant 0$, it follows from (\ref{e:AA})
that $A_+\leqslant A_-$, then  (\ref{e:logA}) implies that
\begin{align*}
F'(z)\leqslant&-\frac14\left(\frac{rc_3\tanh x+(c_3^2-b^2)z\tanh^2x}{1+sz\tanh x}\frac1{A_+}\log_2\frac{1+A_+}{1-A_+}\right.\\
&\left.\quad\quad\quad+\frac{-rc_3\tanh x+(c_3^2-b^2)z\tanh^2x}{1-sz\tanh x}\frac1{A_-}\log_2\frac{1+A_-}{1-A_-}\right)\\
%\leqslant&-\frac14\left(\frac{rc_3\tanh x+(c_3^2-b^2)z\tanh^2x}{1+sz\tanh x}\frac1{A_+}\log_2\frac{1+A_+}{1-A_+}\right.\\
%&\left.\quad\quad\quad+\frac{-rc_3\tanh x+(c_3^2-b^2)z\tanh^2x}{1+sz\tanh x}\frac1{A_-}\log_2\frac{1+A_-}{1-A_-}\right)\\
\leqslant&-\frac14\left(\frac{rc_3\tanh x+(c_3^2-b^2)z\tanh^2x}{1+sz\tanh x}+\frac{-rc_3\tanh x+(c_3^2-b^2)z\tanh^2x}{1+sz\tanh x}\right)\frac1{A_+}\log_2\frac{1+A_+}{1-A_+}\\
=&-\frac12\frac{(c_3^2-b^2)z\tanh^2x}{1+sz\tanh x}\frac1{A_+}\log_2\frac{1+A_+}{1-A_+}\leqslant 0.
\end{align*}

(ii) If $s\tanh x\geqslant 0,rc_3\tanh x\leqslant 0$ and $src_3\leq c_3^2-b^2\leqslant 0$ , we have that
\begin{align*}
F'(z)\leqslant&-\frac14\left(\frac{rc_3\tanh x+(c_3^2-b^2)z\tanh^2x}{1+sz\tanh x}\frac1{A_+}\log_2\frac{1+A_+}{1-A_+}\right.\\
&\left.\quad\quad\quad+\frac{-rc_3\tanh x+(c_3^2-b^2)z\tanh^2x}{1-sz\tanh x}\frac1{A_-}\log_2\frac{1+A_-}{1-A_-}\right)\\
\leqslant &-\frac14\left(\frac{rc_3\tanh x+(c_3^2-b^2)z\tanh^2x}{1+sz\tanh x}+\frac{-rc_3\tanh x+(c_3^2-b^2)z\tanh^2 x}{1-sz\tanh x}\right)\frac1{A_-}\log_2\frac{1+A_-}{1-A_-}\\
=&-\frac12\frac{(-rc_3s+c_3^2-b^2)z\tanh^2x}{(1-sz\tanh x)(1+sz\tanh x)}\frac{1}{A_-}\log_2\frac{1+A_-}{1-A_-}\leqslant 0.
\end{align*}

%(iii) $s\tanh x=0$ and $c_3^2-b^2\geqslant 0$, so $A_{\pm}=H_{\pm}$. Note that $A_+-A_-$ has the same sign as $rc_3\tanh x$ due to  (\ref{e:AA}). Then
%\begin{align*}
%F^{\prime}(z)=&-\frac14\left(\frac{rc_3\tanh x+(c_3^2-b^2)z\tanh^2x}{A_+}\log_2\frac{1+A_+}{1-A_+}\right.\\
%&\left.\quad\quad\quad+\frac{-rc_3\tanh x+(c_3^2-b^2)z\tanh^2x}{A_-}\log_2\frac{1+A_-}{1-A_-}\right)\\
%=&-\frac{rc_3\tanh x}{4}\left(\frac1{A_+}\log_2\frac{1+A_+}{1-A_+}-\frac1{A_-}\log_2\frac{1+A_-}{1-A_-}\right)\\
%&-\frac{(c_3^2-b^2)z\tanh^2x}{4}\left(\frac1{A_+}\log_2\frac{1+A_+}{1-A_+}
%+\frac1{A_-}\log_2\frac{1+A_-}{1-A_-}\right)\\
%\leqslant&0,
%\end{align*}
%We see that $F^{\prime}(z)$ is always decreasing, so the minimum of $F(z)$ is $F(1)$, which is simplified
%to the formula shown in  (\ref{case-a}).

{\it Case (b)} is treated in two subcases, (i) Suppose that $s\tanh x\leqslant 0, rc_3\tanh x\geqslant 0$, and $c_3^2-b^2\geqslant 0$,
(ii) Suppose that $s\tanh x\leq 0,rc_3\tanh x\geqslant 0$ and $src_3\leq c_3^2-b^2\leq 0$. We can solve the problem by the same method as {\it Case (a)}.

{\it Case (c)} If $s=r=0$, then
\begin{align*}
F'(z)= -2H_+^{-1}(c_3^2-b^2)z\tanh^2x\log_2\frac{1+H_+}{1-H_+}.
\end{align*}
Therefore $\min F(z)$ is $F(0)$ according to $c_3^2\leq b^2$ or not. Thus,
the minimum of $F(z)$ on $z\in[0,1]$ is $F(0)$.%given by the same formula in (\ref{case-c}).

{\it Case (d)}. If $s=rc_3$, $b^2=c_3^2$, and $r^2+c_3^2\tanh^x\pm rc_3\tanh x\geq1$, it follows from (\ref{e:AA}) that the first term $F'(z)\geq 0$.

Let $k(z)=\frac{rc_3\tanh x}{H(z)}\log_2\frac{1+sz\tanh x z+H(z)}{1+sz\tanh x z -H(z)}$, where $H(z)=\sqrt{r^2+b^2\tanh^2x+2sz\tanh x }$. Then
\begin{align}
F'(z)\geq&-\frac{rc_3\tanh x}{4H_+(z)}\log_2\frac{1+sz\tanh x +H_+}{1+sz\tanh x -H_+}-\frac{-rc_3\tanh x}{4H_-(z)}\log_2\frac{1-sz\tanh x +H_-}{1-sz\tanh x -H_-}\notag\\
=&-\frac14(k(z)-k(-z)).
\end{align}
As a function of $z$ we have that $H'(z)=\frac{s\tanh x}{H(z)}$, $H''(z)=-\frac{s^2\tanh^2x}{H^3(z)}$ and
\begin{align*}
k'(z)&=H''(z)\log_2\frac{1+sz\tanh x +H(z)}{1+sz\tanh x -H(z)}+\frac{s\tanh x}{H(z)\ln2}\left(\frac{s\tanh x+H'(z)}{1+sz\tanh x +H(s)}-\frac{s\tanh x-H'(z)}{1+sz\tanh x -H(z)}\right)\notag\\
%&=-\frac{s^2\tanh^2x}{H^3(z)}\log_2\frac{1+sz\tanh x +H(z)}{1+sz\tanh x -H(z)}+\frac{2s^2\tanh^2x}{H^2(z)\ln2}\frac{1+sz\tanh x -H^2(z)}{(1+sz\tanh x )^2-H^2(z)}\notag\\
&=-\frac{s^2\tanh^2x}{H^3(z)}\log_2\frac{1+sz\tanh x +H(z)}{1+sz\tanh x -H(z)}+\frac{2s^2\tanh^2x}{H^2(z)\ln2}\frac{1-sz\tanh x -r^2-b^2\tanh^2x}{(1+sz\tanh x )^2-H^2(z)}\leq0,
\end{align*}
the inequality holds because
\begin{align*}
r^2+c_3^2\tanh^2x\pm rc_3\tanh x\geq1.
\end{align*}
Similarly $k'(-z)\leq0$, thus $-\frac{1}{4}(k'(z)+k'(-z))\geq0$, which implies that $F'(z)\geq0$.

Therefore the minimum of $F(z)$ on $z\in[0,1]$ is $F(0)$. %which is given by the formula in (\ref{case-c}).

\bibliographystyle{amsalpha}

\end{document}